\documentclass[journal]{IEEEtran}

\usepackage{graphicx}
\usepackage{cite}
\usepackage{picinpar}
\usepackage{amsmath}
\usepackage{url}
\usepackage[latin1]{inputenc}
\usepackage{colortbl}
\usepackage{soul}
\usepackage{multirow}
\usepackage{pifont}
\usepackage{color}
\usepackage{alltt}
\usepackage[hidelinks]{hyperref}
\usepackage{enumerate}
\usepackage{siunitx}
\usepackage{breakurl}
\usepackage{epstopdf}
\usepackage{pbox}
\usepackage{mathtools}
\usepackage{subcaption}
\usepackage{tabulary}
\usepackage{epsfig}
\usepackage{booktabs}
\usepackage{amssymb}
\usepackage[T1]{fontenc}

\DeclareGraphicsExtensions{.pdf,.png,.jpg,.eps}

\title{Privacy Protection via Joint Real and Reactive Load Shaping in Smart Grids}

\begin{document}

%\title{Real and Reactive Power Based Optimal Privacy Protection in Smart Grid Demand Response}

%\titleheader{2017 IEEE 999999th International Something Conference}

\author{Cihan~Emre~Kement,~\IEEEmembership{Member,~IEEE},~Marija~Ili\'c,~\IEEEmembership{Fellow,~IEEE},~Hakan~Gultekin, Cihan~Tugrul~Cicek, and Bulent~Tavli,~\IEEEmembership{Senior~Member,~IEEE}% <-this % stops a space
\thanks{C. E. Kement and M. Ili\'c are with the Laboratory for Information and Decision Systems (LIDS), Massachusetts Institute of Technology (MIT), Cambridge,
MA, 02139 USA e-mail: \{kement, ilic\} [at] mit [dot] edu.}
\thanks{H. Gultekin is with Sultan Qaboos University, Muscat, Oman e-mail: \{hgultekin\} [at] squ [dot] edu [dot] om.}
\thanks{C. T. Cicek is with Atilim University, Ankara, 06830 Turkey e-mail: \{cihan.cicek\} [at] atilim [dot] edu [dot] tr.}
\thanks{C. E. Kement, H. Gultekin and B. Tavli are with TOBB University of Economics and Technology, Ankara, 06510 Turkey e-mail: \{ckement, hgultekin, btavli\} [at] etu [dot] edu [dot] tr.}
}

\maketitle

\begin{abstract}
Frequent metering of electricity consumption is crucial for demand side management in smart grids. However, metered data can be processed fairly easily by employing well-established nonintrusive appliance load monitoring techniques to infer appliance usage, which reveals information about consumers' private lives. Existing load shaping techniques for privacy primarily focus only on altering metered real power, whereas smart meters collect reactive power consumption data as well for various purposes. This study addresses consumer privacy preservation via load shaping in a demand response scheme, considering both real and reactive power. We build a multi-objective optimization framework that enables us to characterize the interplay between privacy maximization, user cost minimization, and user discomfort minimization objectives. Our results reveal that minimizing information leakage due to a single component, e.g., real power, would suffer from overlooking information leakage due to the other component, e.g., reactive power, causing sub-optimal decisions. In fact, joint shaping of real and reactive power components results in the best possible privacy preservation performance, which leads to more than a twofold increase in privacy in terms of mutual information.
\end{abstract}


\IEEEpeerreviewmaketitle

\section*{Nomenclature}
\addcontentsline{toc}{section}{Nomenclature}
\begin{IEEEdescription}[\IEEEusemathlabelsep\IEEEsetlabelwidth{$V_1,V_2,V_3$}]
\item[\textbf{Indices and sets}]
\item[$a$] Index of appliances
\item[$i$] Index of objectives
\item[$as$] Index of appliance scenarios
\item[$rs$] Index of renewable scenarios
\item[$t$, $\tau$] Indices of time slots
\item[$A^{ts}$] Set of time-shiftable appliances
\item[]
\item[\textbf{Parameters}]
\item[$\alpha_{a}$] Operation window start of appliance $a$
\item[$\beta_{a}$] Operation window end of appliance $a$
\item[$\gamma_i$] Weight of objective $O_i$
\item[$\Delta^{t}$] Duration of one time slot
\item[$\eta^{cp}$] Charge efficiency of the battery
\item[$\eta^{dp}$] Discharge efficiency of the battery
\item[$\eta^{cq}$] Charge efficiency of the capacitor
\item[$\eta^{dq}$] Discharge efficiency of the capacitor
\item[$\phi_{a,t}$] Penalty cost of appliance $a$ for operating at time slot $t$
\item[$\rho_{as}$] Probability of appliance scenario $as$
\item[$\rho_{rs}$] Probability of renewable scenario $rs$
\item[$c^p_t$] Cost of real power at time slot $t$ (\$/kWh)
\item[$E_{a}$] Amount of energy that appliance $a$ has to spend to complete its operation (kWh)
\item[$E^{bi}$] Initial energy stored in the battery (kWh)
\item[$E^{bmax}$] Maximum energy that can be stored in the battery (kWh)
\item[$E^{ci}$] Initial reactive energy stored in the capacitor (kvarh)
\item[$E^{cmax}$] Maximum reactive energy that can be stored in the capacitor (kvarh)
\item[$O^{\ast}_{i}$] Optimal value of $O_i$
\item[$P^g_{rs,t}$] Real power generated by the PV generator at time slot $t$ in scenario $rs$ (kW)
\item[$P^{max}$] Load capacity of the house (kW)
\item[$P^{max}_{a}$] Maximum real power that appliance $a$ can draw during its operation (kW)
\item[$P^{min}_{a}$] Minimum real power that appliance $a$ can draw during its operation (kW)
\item[$PF_{a}$] Power factor of appliance $a$
\item[$P^{od}_{as,t}$] Real power used by on-demand appliances at time slot $t$ in scenario $as$ (kW)
\item[$P^{sc}_{t}$] Real power used by safety-critical appliances at time slot $t$ (kW)
\item[$Q^{od}_{as,t}$] Reactive power used by on-demand appliances at time slot $t$ in scenario $as$ (kvar)
\item[$Q^{sc}_{t}$] Reactive power used by safety-critical appliances at time slot $t$ (kvar)
\item[$R^{cbmax}$] Maximum charge rate of the battery (kW)
\item[$R^{dbmax}$] Maximum discharge rate of the battery (kW)
\item[$R^{ccmax}$] Maximum charge rate of the capacitor (kvar)
\item[$R^{dcmax}$] Maximum discharge rate of the capacitor (kvar)
\item[]
\item[\textbf{Variables}]
\item[$O_i$] Value of the objective $i$
\item[$p^m_{t}$] Metered real power at time slot $t$ (kW)
\item[$q^m_{t}$] Metered reactive power at time slot $t$ (kW)
\item[$p^{ca}_{a,t}$] Real power consumed by appliance $a$ at time slot $t$ (kW)
\item[$q^{ca}_{a,t}$] Reactive power consumed by appliance $a$ at time slot $t$ (kvar)
\item[$p^{cb}_{t}$] Real power charged into the battery at time slot $t$ (kW)
\item[$p^{db}_{t}$] Real power discharged from the battery at time slot $t$ (kW)
\item[$q^{cc}_{t}$] Reactive power charged into the capacitor at time slot $t$ (kvar)
\item[$q^{dc}_{t}$] Reactive power discharged from the capacitor at time slot $t$ (kvar)
\item[$v_{rs,t}$] Real power drawn from the PV generator at time slot $t$ in scenario $rs$ (kW)
\item[$y_{a,t}$] A binary variable that takes the value 1 if time-shiftable appliance $a$ operates at time slot $t$
\end{IEEEdescription}

\section{Introduction}
\label{sec:Introduction}

\IEEEPARstart{S}{mart} metering is one of the essential constituents of the smart grid (SG). Frequent measurements collected from smart meters are used for accurate and personalized billing services, detecting outages and electricity theft, load forecasting, and demand side management (DSM), among others~\cite{mcdaniel2009security}. However, smart meter data can also be exploited by adversaries to violate the privacy of consumers~\cite{marmol2012donot}.

The generic term for methods used to extract appliance or end-use data from aggregated household meter data is known as nonintrusive appliance load monitoring (NIALM). NIALM methods, typically, detect edges on time-series meter data and employ techniques, such as cluster analysis, to map the change in the metered data to an appliance or end-use~\cite{zoha2012non}. The change in the metered power and other transient and steady-state properties such as duration and periodicity are used as features in the analysis.

A plethora of methods have been proposed to mitigate the privacy problem induced by smart metering and NIALM. These can be grouped into five categories: (i) adding noise to metered data to achieve differential privacy, (ii) using homomorphic encryption techniques to hide sensitive information within metered data, (iii) using pseudonyms instead of consumer identification, (iv) reducing metering frequency, and (v) shaping metered load. Among these methods, load shaping (LS) (also called demand shaping -- DS) is one of the most promising ones in terms of simplicity, efficiency, and applicability\cite{giaconi2018privacy}.

Many LS algorithms and techniques have been proposed in the literature for shaping the real power ($P$) consumed in a household to avoid unveiling appliance-specific signatures~\cite{kement2017comparative}. These techniques utilize household amenities such as rechargeable batteries (RBs), renewable energy sources (RESs), and appliances like plug-in electric vehicles (PEVs) whose runtime and/or power consumption can be shifted for real power shaping.

%Numerous LS methods have been proposed in the literature \cite{kement2017comparative}, most of which are based on algorithms, heuristics and game theory. These methods focus on shaping the metered real power ($P$) consumed by the customer in such a way that it does not give away appliance-specific signatures. For shaping the real power, they use household amenities such as rechargeable batteries (RB), renewable energy sources (RES) as well as some appliances runtime and/or power consumption of which can be shifted (\emph{e.g.} plug-in electric vehicles).

\begin{figure}
  \begin{center}
    \includegraphics[width = \linewidth]{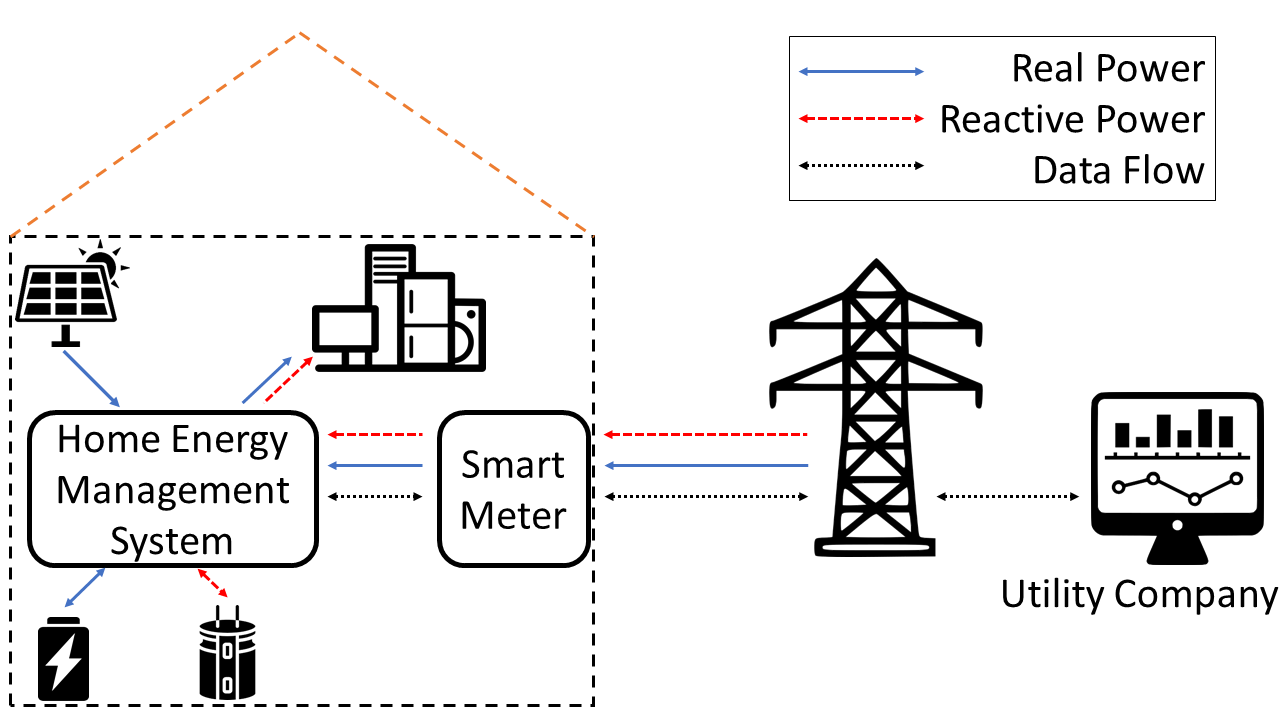}
    \caption{Real and reactive power metering in a smart home.}
    \label{fig:fig1}
  \end{center}
\end{figure}

Although LS-based privacy preservation literature is extensive, an important consideration has been left unaddressed -- smart meters do not measure solely the real power. In fact, they measure instantaneous voltage and current, hence, the complex power ($S$), which has both real ($P$) and reactive ($Q$) power components. Since consumers are usually billed based on their real power consumption, the effects of reactive power on privacy are, mostly, overlooked in the literature. However, metering reactive power is also important for the supply side since having a high reactive load decreases the power factor ($P/\left|S\right|$), hence the grid's efficiency. Therefore, utility companies (UCs) keep track of the reactive power usage as well, as illustrated in Fig. \ref{fig:fig1}.

%Although the number of these LS methods are extensive, they miss a key point: smart meters do not measure only the real power. They measure instantaneous voltage and current, and hence the complex power ($S$), which has both real ($P$) and reactive ($Q$) power components. Since consumers are usually billed based on the real power consumption, the reactive power is often overlooked by the smart grid privacy research community. However, reactive power is also a matter of importance for the supply side, since having a high reactive load decreases the power factor ($P/\left|S\right|$) and hence the efficiency of the system. Therefore, utility companies (UC) also keep track of the reactive power demand as depicted in Fig. \ref{fig:fig1}.

Just like metered real power, metered reactive power also contains signatures of individual appliances~\cite{fan2017privacy}. In fact, temporal variations in metered reactive power are one of the key features exploited by NIALM methods~\cite{zeifman2011nonintrusive}. Therefore, to decrease the amount of information that can be gathered from smart meter data, not only the real power but also the reactive power must be considered. However, joint shaping of real and reactive power for privacy preservation is left unaddressed in the literature. To fill this gap, in this study, we investigate the extent of privacy vulnerabilities of real and reactive power metering as well as the effectiveness of countermeasures for privacy preservation.
%In this study, we present a novel multi-objective optimization problem which maximizes the privacy of the consumer (by shaping $P$, $Q$ or both) while also considering the cost and the comfort of the consumer. We employed real and reactive power measurements from the Ampds2 dataset~\cite{makonin2016electricity} in our analyses to show that hiding only $P$ or only $Q$ is not enough to ensure privacy. 
Our main contributions are enumerated as follows:
\begin{enumerate}
  \item To the best of our knowledge, this is the first study that considers the impacts of shaping both real and reactive loads simultaneously for consumer privacy preservation in SGs.
  \item We develop a novel multi-objective mixed-integer programming (MIP) model for finding the optimum schedule of real and reactive power consumption of a house to maximize privacy while minimizing cost and discomfort. Our formulation employs a minimax goal programming approach to guarantee Pareto efficiency and enables fair prioritization of the aforementioned objectives.
  \item We show that shaping only one of the load components is insufficient to attain the best possible privacy performance. Our computational study, in which we compare different cases such as only real power based LS, only reactive power based LS, and both real and reactive power based LS, confirm that the best possible privacy preservation can only be achieved by joint consideration of real and reactive power.
\end{enumerate}

The rest of the paper is organized as follows: Section~\ref{sec:Related} presents a literature review on LS-based privacy protection in SGs. Our system model and multi-objective MIP formulation are elaborated in Sections~\ref{sec:SystemModel}~and~\ref{sec:ProblemFormulation}, respectively. Results of our analysis are presented in Section~\ref{sec:Results}. Concluding remarks are provided in Section~\ref{sec:Conclusion}.

\section{Related Work}
\label{sec:Related}

The literature on SG privacy is extensive, therefore, we refer the readers to the excellent survey/overview papers on SG privacy~\cite{kumar2019smart,giaconi2018privacy,ashgar2017smart}. Nevertheless, in this section, we provide a concise review of LS-based privacy preservation studies in SGs.

Kalogridis~\textit{et al.}~\cite{kalogridis2010privacy} (Best Effort -- BE -- algorithm), McLaughlin~\textit{et al.}~\cite{mclaughlin2011protecting} (Non Intrusive Load Leveling -- NILL -- algorithm), Ge and Zhu~\cite{ge2013preserving} (Tolerable Deviation -- TD -- algorithm), and Yang~\textit{et al.}~\cite{yang2012minimizing} (a family of stepping algorithms) proposed heuristic algorithms which laid the foundations for LS-based privacy preservation in SGs. In these studies, to reduce (ideally, to eradicate) information leakage, temporal characteristics of metered real load are shaped with the help of RBs, RESs, and/or time/power shiftable appliances.

Building upon the initial solution approaches, alternative/improved/complementary solutions for LS-based privacy preservation employing various amenities have been proposed. Zhao~\textit{et al.}~\cite{zhao2014achieving} used RBs to impose random variations on the metered load to ensure differential privacy. Egarter~\textit{et al.}~\cite{egarter2014load} utilized shiftable appliances instead of RBs to shape the load. Giaconi~\textit{et al.}~\cite{giaconi2018privacy} employed RBs and RESs to create a privacy preserving energy management policy. Chen~\textit{et al.}~\cite{chen2015preventing} proposed the use of thermal storage, such as water heaters, instead of RBs to provide privacy preservation. Reinhardt~\textit{et al.}~\cite{reinhardt2015worried} proposed a method for privacy preservation by utilizing photovoltaic (PV) generators. Liu~\textit{et al.}~\cite{liu2017information} explored SG privacy in the presence of multiple RBs in a cascaded fashion. Sun~\textit{et al.}~\cite{sun2018smart} proposed utilizing PEVs and household appliances such as HVAC (Heating, Ventilating, and Air Conditioning) as energy storage in shaping the load for privacy preservation. Moon~\textit{et al.}~\cite{moon2015usages} presented an optimization framework considering both privacy and cost by using an RB. Liu and Cheng~\cite{liu2017achieving} proposed an optimization model by considering cost, privacy, and comfort with the help of shiftable appliances and RBs. Tan~\textit{et al.}~\cite{tan2017privacy} investigated optimal privacy-cost trade-off with the help of household RBs. Isikman~\textit{et al.}~\cite{isikman2016power} optimized the privacy and power usage (utility) of consumers with the help of RBs and RESs.

Different aspects of the LS-based privacy preservation problem in SGs have been investigated in the literature. Erdemir~\textit{et al.}~\cite{erdemir2019privacy} investigated optimal RB policies by reformulating the privacy optimization as a Markov decision process. Chen and Wu~\cite{chen2013residential} and Yang~\textit{et al.}~\cite{yang2015cost} proposed online algorithms for privacy preservation using RBs and shiftable appliances. Koo~\textit{et al.}~\cite{koo2017rl} proposed a learning-based LS scheme to hide both high and low-frequency load signatures for privacy preservation. Hossain~\textit{et al.}~\cite{hossain2019enhanced} and Natgunanathan~\textit{et al.}~\cite{natgunanathan2019progressive} proposed online and offline heuristic methods to mitigate the problem of preserving privacy in case of prolonged high or low load periods and finite capacity RBs. Ghasemkhani~\textit{et al.}~\cite{ghasemkhani2019learning} modeled the privacy-protecting behavior of consumers in a demand response scheme as a Stackelberg game. They proposed a reinforcement learning algorithm to obtain the optimal strategy of a privacy-aware consumer. Shateri~\textit{et al.}~\cite{shateri2020privacy} proposed a model-free deep reinforcement learning algorithm run by a household controller to ensure the privacy of consumers. Giaconi~\textit{et al.}~\cite{giaconi2020privacy} proposed several energy management policies which utilize LS for attaining privacy-cost balance.

All of the aforementioned studies focused on the minimization of information leakage due to real power only. Fan~\textit{et al.}~\cite{fan2017privacy} is the only study in the literature that revealed the potential privacy risks involved with reactive power. In particular, to mask the true reactive power demand, a solution based on the use of a capacitor to store and provide reactive power in a controlled manner is proposed. The proposed solution, called reactive power obfuscation, is fairly successful in masking the ON/OFF events of home appliances.

All studies in the literature on LS-based privacy preservation considered only real power or only reactive power. However, our results reveal that real and reactive powers must be shaped simultaneously for the best privacy preservation performance. Indeed, we show that shaping real power without shaping the reactive power (and vice versa) leads to significantly higher information leakage when compared to joint privacy preservation. 

\section{System Model}
\label{sec:SystemModel}

In this section, we present the models that our optimization formulation is built upon. In the following subsections we elaborate on the demand response model, battery model, PV energy model, and appliance models.

\subsection{Demand Response and Pricing}

We consider a smart-metered house with various appliances, a household battery and a capacitor for real and reactive power shaping, respectively, a PV generator, and a home energy management system (HEMS) for scheduling the appliances as illustrated in Fig.~\ref{fig:fig1}. We assume a price-based demand response (DR) scheme \cite{albadi2007demand} is in effect where the UC provides the day-ahead time-of-use (ToU) pricing information to the consumer, and the HEMS determines the optimal schedule of the appliances with respect to cost, comfort, and privacy priorities of the consumer.

%The appliances are classified into 3 groups. The first group contains time-shiftable appliances, which have certain operation windows within which they must complete their operation. The second group contains power-and-time-shiftable appliances which have maximum and minimum power levels within which they can operate (as well as time-shiftability). The third group containts other appliances that cannot be scheduled beforehand as they are either safety-critical appliances or on-demand appliances. 

%Note that, in this model, the HEMS is also tasked with estimating the power generated by the PV and the power requirements of the on-demand appliances. We assume that both PV generation (i.e., renewable scenarios -- $rs$) and on-demand appliance scenarios ($as$) follow probability distributions that are known beforehand. PV generation varies depending on weather conditions whereas on-demand appliance usage varies with respect to consumer preferences.

\subsection{Household Battery and Capacitor}

In this study, we assume that a household battery with a maximum capacity of $E^{bmax}$ in kWh, and charging and discharging efficiencies of $\eta^{cp}$ and $\eta^{dp}$, respectively, and a capacitor with a maximum capacity of $E^{cmax}$ in kvar, and charging and discharging efficiencies of $\eta^{cq}$ and $\eta^{dq}$, respectively, are present in the house. Since the planning horizon in our model is 24 hours, long-term aspects of having a household battery, such as the investment cost or life-cycle degradation, are neglected.

\subsection{Solar Irradiance Scenarios and PV Energy Model}
\label{sec:solar}

We use the global tilted irradiance (GTI) data collected in the Alderville region of Canada~\cite{canada_2020} to account for the PV energy generation. In this data set, global horizontal irradiance (GHI) and GTI data are available through four different days with different weather conditions. The irradiance data was measured once at each ms and averaged over minutely intervals. The dataset provides measurements from 24 individual sensors located in close proximity.

Based on this GTI data, four solar irradiance scenarios are generated. Since a historical dataset to infer the probability of weather conditions is not available, we assume that each scenario has an equal probability of $\rho_{rs}$. Consequently, the power generated by the PV generator under each scenario is obtained by using the irradiance-to-power conversion function~\cite{mazidi2014Integrated} as:
\begin{equation}\label{eq:conversion}
	P^g_{rs,t} = \eta_{pv} \cdot \S_{pv} \cdot \textrm{GTI}_{rs,t}, \quad \forall rs, t,
\end{equation}
where $\S_{pv}$ and $\eta_{pv}$ represent the solar panel area in $\textrm{m}^2$ and the efficiency of solar panels, respectively.

\subsection{Appliance Scenarios and Modeling}
\label{subsec:Appliance}

Appliances are categorized into three groups according to their level of programmability. The first group contains on-demand and safety-critical appliances which cannot be scheduled temporally and whose power consumption profiles cannot be changed (i.e., non-shiftable appliances). Examples of on-demand and safety-critical appliances are TVs and CCTV systems, respectively. The second group contains appliances whose operation can be delayed and/or interrupted as needed, yet, their power consumption profiles cannot be altered (i.e., time-shiftable appliances). Examples of such appliances are clothes washers and dryers. The third group contains appliances whose power consumption levels can also be changed; thus, they can be scheduled temporally and spatially (i.e., power-and-time-shiftable appliances).
%The only difference between time-shiftable and power-and-time-shiftable appliances is that the power usage of time-shiftable appliances cannot be altered.
Such appliances can be used somewhat similar to a household battery, yet, they need to satisfy certain additional constraints. For example, PEVs are power-and-time-shiftable appliances and can be charged/discharged according to consumers' needs. However, they have to be charged enough at a certain time (e.g., at 8:00 a.m.) to be practical. Another example is HVAC systems, which are power-and-time-shiftable. However, they must keep the temperature within a certain range (e.g., $20$-$24^{\circ}C$).

There exist only a few appliance-level datasets with real and reactive power measurements available as presented in~Table~\ref{tab:datasets}. Among these datasets, the measurement duration of an appliance in ACS-F2~\cite{ridi2014acs} is two hours. Hence, the dataset does not include information about the daily usage of appliances. On the other hand, most of the measured appliances in Ampds2 dataset~\cite{makonin2016electricity} and iAWE dataset~\cite{batra2013s} overlap, and the measurement duration in iAWE dataset is shorter than that of Ampds2 and it varies with the appliance. Therefore, in this study we use the appliance power measurement data from Ampds2 dataset. %Note that all the appliances we considered in our scenarios have real power components.

\begin{table}[!htb]
\centering \caption{Datasets with appliance level real and reactive power measurements.}
    \begin{tabulary}{7.2cm}{llll}
    \toprule
    Dataset&Duration&Period&\# of Appliances\\
    \cmidrule{1-4}
    Ampds2&$2$ years&$1$ minute& $10$\\
    ACS-F2&$2$ hours&$10$ seconds& $15$\\
    iAWE&$24-102$ days&$1$ second&$10$\\\bottomrule
\end{tabulary}\label{tab:datasets}
\end{table}

We embed real and reactive power usages of safety-critical appliances ($P^{sc}_t$, $Q^{sc}_t$) into the model as parameters as stated in~\eqref{eq:cons12}. The on-demand power usage scenarios ($P^{od}_{as,t}$, $Q^{od}_{as,t}$) are obtained by using k-means clustering~\cite{saghezchi2014Game} over the two-year consumption data of the on-demand appliances in Ampds2 dataset.
%Time series power usage data of non-shiftable appliances was directly added to the power balance constraint \eqref{eq:cons12}. Similarly, the time series data of time-shiftable appliances which can be delayed but cannot be interrupted was kept intact. The constraint XXX ensures that once these devices start their operation, their schedule is not interrupted. Interruptable time-shiftable appliances were broken into sub-operations and they are treated as individual non-interruptable appliances with sequential operation constraint (see XXX).
Power-and-time-shiftable appliances are modeled to complete their operations within their operation windows [$\alpha_a$-$\beta_a$] and their power usages are within $P^{min}_a$ and $P^{max}_a$.
%{\color{red} The real and reactive power usage of safety-critical appliances are assumed to be known a priori and given as parameters to the problem. The time-shiftable appliances are modeled similar to the power-and-time-shiftable appliances, except that their power usage cannot be shifted. The on-demand power usage scenarios are obtained by reducing the two-year consumption data of the on-demand appliances in Ampds2 dataset with k-means clustering \cite{saghezchi2014Game}.

To obtain the appliance scenarios ($as$), we first aggregate each daily on-demand electricity usage data and take each day as a scenario. As a result, we obtain 730 scenarios. Then, the k-means algorithm with $k=10$ is applied to these 730 scenarios to reduce the number of scenarios to 10 for having a reasonably sized problem. Probability of each on-demand appliance scenario, $\rho_{as}$, is set to the ratio between the associated cluster size and the total number of scenarios.

\section{Problem Formulation}
\label{sec:ProblemFormulation}

\subsection{Objectives}
\label{subsec:objetives}

Our objectives are maximization of consumer privacy by minimizing the information carried within both real and reactive metered loads, minimization of consumer cost and minimization of consumer discomfort. Indeed, there are inherent trade-offs among these three objectives. In fact, accounting for cost and discomfort are important considerations in the holistic characterization of privacy preserving SG operation.

\subsubsection{Maximizing Privacy by Shaping Real and Reactive Power}
\label{subsubsec:maxprivacy}

We define the privacy objective as a generic function, $F$, of the metered real ($p^m_{t}$) and reactive ($q^m_{t}$) power of the house. $F$ can be formulated depending on the method that will be used for preserving privacy. In~\cite{kement2017comparative}, comparative performance evaluations of well-known LS-based privacy preservation techniques have been presented, and it has been shown that the BE technique~\cite{kalogridis2010privacy} is one of the best performing privacy preservation techniques. Therefore, we adopt the BE technique, which has originally been proposed for real power based privacy preservation. We generalize the idea behind BE technique to encompass both $p^m_{t}$ and $q^m_{t}$ as
\begin{equation} \label{eq:o0}
    F \triangleq \sum_{t=2}^{T} \big(\left|p^m_{t}-p^m_{t-1}\right|+\left|q^m_{t}-q^m_{t-1}\right|\big).
\end{equation}

Note that it is also possible to adopt other LS-based load shaping techniques (which have originally been designed for real power) to come up with a solution to address both real and reactive power privacy. However, such an endeavor is beyond the scope of our study.

%It is not possible to assess the impact of shaping $p^m_{t}$ and $q^m_{t}$ individually on privacy preservation by utilizing~\eqref{eq:o0} because temporal variations of metered real and reactive loads have the same impact on privacy according to~\eqref{eq:o0}. 
Since shaping $p^m_{t}$ and $q^m_{t}$ are expected to have temporal variations in metered real and reactive loads, we divide~\eqref{eq:o0} into two separate privacy objectives so that we can weight them accordingly to explore different cases. Mathematical expressions of privacy objectives for real ($O_1$) and reactive load ($O_2$) are presented in~\eqref{eq:o1}--\eqref{eq:o22}, where non-negative variables $d^1_t$, $d^2_t$, $d^3_t$, and $d^4_t$ are used for the linearization of the absolute values in \eqref{eq:o0}.

\begin{align} 
    &O_1 = \sum_{t=2}^{T} \left[ d^1_t + d^2_t +\epsilon \left( p^{cb}_t + p^{db}_t + q^{cc}_t + q^{dc}_t \right) \right] \label{eq:o1}, \\
    &d^1_t - d^2_t = p^m_{t}-p^m_{t-1}, \quad \forall t\geq 2 \label{eq:o11}, \\
    &O_2 = \sum_{t=2}^{T} \left[ d^3_t + d^4_t + \epsilon \left( p^{cb}_t + p^{db}_t + q^{cc}_t + q^{dc}_t \right) \right] \label{eq:o2}, \\
    &d^3_t - d^4_t = q^m_{t}-q^m_{t-1}, \quad \forall t\geq 2 \label{eq:o22}.
\end{align}

In particular, when $p^m_{t} - p^m_{t-1} \geq 0$, the minimization of $O_1$ forces $d^1_t$ to be $p^m_{t} - p^m_{t-1}$ and $d^2_t$ to be $0$. Likewise, when $p^m_{t} - p^m_{t-1} < 0$, $d^1_t$ is set to $0$ and $d^2_t$ is set to $p^m_{t-1} - p^m_{t}$. In either case, $O_1$ is increased by $\left| p^m_{t} - p^m_{t-1} \right|$. A similar linearization follows in modeling $O_2$. Moreover, charging/discharging variables ($p^{cb}_t$, $p^{db}_t$, $q^{cc}_t$, and $q^{dc}_t$) are included in the objectives to prevent the occurrence of simultaneous charging/discharging events in the same time slot. Since the exclusion of charge/discharge events is not the main objective, these variables are multiplied by a sufficiently small penalty coefficient, $\epsilon > 0$.

\subsubsection{Minimizing Monetary Cost}
\label{subsubsec:mincost}

In a traditional SG, reactive power usage is not billed to residential consumers. Therefore, the monetary cost is based only on the real power. In particular, the monetary cost ($O_3$) is defined as the product of metered real energy ($\Delta^t \cdot p^m_{t}$) and the price of electricity in time slot $t$ ($c^p_t$), and stated as
\begin{equation} \label{eq:o3}
    O_3 = \Delta^t \sum_{t=1}^{T} c^{p}_t p^m_{t}.
\end{equation}
%Since the reactive power is not billed to residential consumers, its cost is not added to the formulation.

Note that in~\eqref{eq:o3}, no penalty function is added for $p^{cb}_t$, $p^{db}_t$, $q^{cc}_t$, and $q^{dc}_t$ since this penalty is already implied when~\eqref{eq:o3} is minimized along with the constraints~\eqref{eq:cons12} and~\eqref{eq:cons14}. The charge/discharge inefficiencies of the battery and the capacitor given in~\eqref{eq:cons12} and~\eqref{eq:cons14} result in higher $p^m_{t}$ if the battery or the capacitor is charged and discharged in the same time slot. Since~\eqref{eq:o3} is minimizing $p^m_{t}$, the optimal solution of this objective yields mutually exclusive charge/discharge events.
%Although currently consumers are not charged for their reactive power consumption, high reactive loads induce inefficiency to the grid, which is indirectly reflected on the pricing by the UC. In addition, incurring a cost on reactive power would also eliminate the possibility of unnecessary reactive power consumption in the optimal solution. We assume a constant and relatively small cost $c^{q}$ for the reactive power usage.

\subsubsection{Minimizing Discomfort}
\label{subsubsec:mindiscomfort}

LS causes a certain discomfort to a consumer if appliance operations are shifted to later time slots. We model the discomfort ($O_4$) by defining an exponentially increasing penalty coefficient ($\phi_{a,t} = \left(t-\alpha_a\right)^2/E_a, \quad \forall a, t \in [\alpha_{a},\beta_{a}]$) \cite{Mohsenian-Rad2010OptimalResidential} for each appliance. As a result, $O_4$ is defined as
\begin{equation} \label{eq:o4}
    O_4=\sum_{a=1}^{A}\sum_{t=1}^{T} \left[ \phi_{a,t} p^{ca}_{a,t} + \epsilon \left( p^{cb}_t + p^{db}_t + q^{cc}_t + q^{dc}_t \right) \right],
\end{equation}

\noindent where charging/discharging activities are penalized as defined in \eqref{eq:o1} and \eqref{eq:o2} for mutual exclusion.

Note that we do not need to penalize the reactive power usage of the appliances in~\eqref{eq:o4} because all the appliance usages in our setup already have real power components. In other words, there are no purely reactive loads present in the Ampds2 dataset. Therefore, consumer comfort can be solely measured in terms of the real power usage. If there were purely reactive loads in the household environment, it would be necessary to include $q^{ca}_{a,t}$ into this objective. 

\subsection{Constraints}
\label{sec:constraints}

We can divide the constraints into three categories: (i) appliance power constraints, (ii) power balance constraints, and (iii) battery/capacitor constraints. All three categories have additional constraints to account for when compared to the classical LS-based privacy preservation studies in the literature due to the inclusion of reactive power.

\subsubsection{Appliance Power Constraints}
\label{subsubsec:applianceconstraints}

Constraint~\eqref{eq:cons8} makes sure that the real power used by appliance $a$ is $0$ outside its operation window
\begin{align}
    &p^{ca}_{a,t}=0, \quad  \forall a, \forall t \notin [\alpha_{a},\beta_{a}]. \label{eq:cons8}
\end{align}
Constraint~\eqref{eq:cons9} guarantees that power-and-time-shiftable appliances run within their power limits
\begin{align}
    &P^{min}_{a}\leq p^{ca}_{a,t}\leq P^{max}_{a}, \quad  \forall a,t. \label{eq:cons9}
\end{align}
Constraint~\eqref{eq:cons10} correctly determines the power usage of time-shiftable appliances
\begin{align}
    &p^{ca}_{a,t}=y_{a,t} \cdot P^{max}_{a}, \quad  \forall t, \forall a \in A^{ts}. \label{eq:cons10}
\end{align}
In particular, when appliance $a$ runs in time slot $t$, the corresponding binary variable $y_{a,t}$ is set to $1$ and $p^{ca}_{a,t} = P^{max}_a$. Otherwise, $y_{a,t}$ is set to 0 and $p^{ca}_{a,t} = 0$. 

Constraint~\eqref{eq:cons11} ensures that each appliance consumes the total energy required to complete its operation
\begin{align}
    &\Delta^t \cdot \sum_{t=1}^{T}p^{ca}_{a,t} = E_{a}, \quad \forall a. \label{eq:cons11}
\end{align}
Constraint~\eqref{eq:cons16} states that real and reactive power consumption of an appliance is proportional to its power factor
\begin{align}
    &q^{ca}_{a,t} = \tan(\arccos(PF_{a})) \cdot p^{ca}_{a,t}, \quad  \forall a,t. \label{eq:cons16}
\end{align}

%In~\eqref{eq:cons16}, we assumed that all appliances have a constant power factor. However, this may not be the case for some appliances with multiple components. Yet, these components can, still, be modeled as individual appliances whose operations are dependent on each other. Thus, their $P$-$Q$ relations can be modeled with a linear relationship as in~\eqref{eq:cons16}.

\subsubsection{Power Balance Constraints}
\label{subsubsec:pbconstraints}

%Constraints~\eqref{eq:cons12}-\eqref{eq:cons13} and \eqref{eq:cons14} are the real and reactive power balance constraints, respectively
Constraints~\eqref{eq:cons12}-\eqref{eq:cons14} are the real and reactive power balance constraints
\begin{align}
    p^m_{t} = &\sum_{a=1}^{A}p^{ca}_{a,t} + P^{sc}_t + \sum_{as}\rho_{as} \cdot P^{od}_{as,t} + \notag \\
    &p^{cb}_{t}\big/\eta^{cp} - p^{db}_{t}\cdot\eta^{dp} - \sum_{rs}\rho_{rs} \cdot v_{rs,t}, \quad \forall t, \label{eq:cons12} \\
    v_{rs,t} \leq& P^g_{rs,t}, \quad  \forall rs, t, \label{eq:cons13} \\
    q^m_{t} = &\sum_{a=1}^{A}q^{ca}_{a,t} + Q^{sc}_t + \sum_{as}\rho_{as} \cdot Q^{od}_{as,t} + \notag \\
    &q^{cc}_{t}\big/\eta^{cq} - q^{dc}_{t}\cdot\eta^{dq}, \quad \forall t. \label{eq:cons14}
\end{align}
More precisely, the real power measured by the smart meter in time slot $t$ is equal to the sum of real power consumed by all three types of appliances, the real power charged into the battery minus the real power discharged from the battery and drawn from the PV generator. Utilized PV power cannot be greater than the PV generation for each scenario as ensured in \eqref{eq:cons13}. The reactive power measured by the smart meter is determined similarly with~\eqref{eq:cons14} except the PV generation term. Lastly, the real power demand of the household is bounded as
\begin{align}
    p^{m}_{t} \leq& P^{max}, \quad \forall t. \label{eq:cons15}
\end{align}

\subsubsection{Battery and Capacitor Constraints}
\label{subsubsec:batcapconstraints}

Constraints~\eqref{eq:cons17} and \eqref{eq:cons18} make sure that capacities of the battery and the capacitor are not exceeded in any time slot $\tau$, respectively,
\begin{align}
    &0 \leq E^{bi} + \sum_{t=1}^{\tau}\Delta^t \cdot p^{cb}_{t} - \sum_{t=1}^{\tau}\Delta^t \cdot p^{db}_{t} \leq E^{bmax}, \forall \tau, \label{eq:cons17} \\
    &0 \leq E^{ci} + \sum_{t=1}^{\tau}\Delta^t \cdot q^{cc}_{t} - \sum_{t=1}^{\tau}\Delta^t \cdot q^{dc}_{t} \leq E^{cmax}, \forall \tau. \label{eq:cons18}
\end{align}

Constraints~\eqref{eq:cons19} and~\eqref{eq:cons20} limit the amount of real power charged into or discharged from the battery in each time slot, respectively,
\begin{align}
    &p^{cb}_{t} \leq R^{cbmax}, \quad \forall t, \label{eq:cons19} \\
    &p^{db}_{t} \leq R^{dbmax}, \quad \forall t. \label{eq:cons20}
\end{align}

Similarly, constraints~\eqref{eq:cons22} and~\eqref{eq:cons23} bound the rate at which reactive power can be stored and provided by the capacitor in each time slot, respectively,
\begin{align}
    &q^{cc}_{t} \leq R^{ccmax}, \quad \forall t, \label{eq:cons22} \\
    &q^{dc}_{t} \leq R^{dcmax}, \quad \forall t. \label{eq:cons23}
\end{align}

Constraints~\eqref{eq:cons21} and~\eqref{eq:cons24} guarantee that the amount of real and reactive power stored at the beginning and at the end of the day are the same, respectively,
\begin{align}
    &\sum_{t=1}^{T}p^{cb}_{t} = \sum_{t=1}^{T}p^{db}_{t}, \label{eq:cons21} \\
    &\sum_{t=1}^{T}q^{cc}_{t} = \sum_{t=1}^{T}q^{dc}_{t}. \label{eq:cons24}
\end{align}

\subsection{Multi-Objective Optimization Model}
\label{subsec:moom}

%We use a \emph{minimax goal programming}~\cite{kement2020holistic} approach to model the optimization problem as expressed in~\eqref{eq:mip0} and~\eqref{mip:cons27}
We use a \emph{minimax goal programming}~\cite{kement2020holistic} approach to formulate the optimization problem. Our goal is to minimize the maximum deviation from each objective's optimal objective value, which is obtained when the model is solved by ignoring all other objectives. In particular, we first find the stand-alone optimal value of each objective ($O_i^\ast$) by solving the following MIP formulation:
\begin{align}
    \nonumber O_i^\ast = \min & \ O_i\\
    \textrm{subject to} & \ \eqref{eq:cons8}~-~\eqref{eq:cons24}, \label{eq:cons_singular1} \\
    & \ \eqref{eq:o1}, \eqref{eq:o11} & \textrm{ if } i = 1, && \label{eq:cons_singular2} \\
    & \ \eqref{eq:o2}, \eqref{eq:o22} & \textrm{ if } i = 2, && \label{eq:cons_singular3} \\
    & \ \eqref{eq:o3} & \textrm{ if } i = 3, && \label{eq:cons_singular4} \\
    & \ \eqref{eq:o4} & \textrm{ if } i = 4. && \label{eq:cons_singular5}
\end{align}

Then, using $O_i^\ast$ values and assigning a weight ($\gamma_i$) to objective $i$, we solve the following minimax MIP formulation:
\begin{align}
    \nonumber \min & \max_{i=1,2,3,4} \left\{ \gamma_i \dfrac{O_{i}-O_{i}^{\ast}}{O_{i}^{\ast}} \right\} \\ 
    \textrm{subject to} & \quad  \eqref{eq:o1}~-~\eqref{eq:cons24}.\nonumber
\end{align}

Note that the objective function in the above formulation is not linear. Therefore, we introduce a new auxiliary variable $Z$ and define the final MIP formulation as
\begin{align}
    \nonumber \min & \quad Z\\ 
    \textrm{subject to} & \quad Z \geq \gamma_{i} \cdot \frac{O_{i}-O_{i}^{\ast}}{O_{i}^{\ast}}, \quad \forall i, \label{mip:cons27} \\ 
    &\quad \eqref{eq:o1}~-~\eqref{eq:cons24}.\nonumber
\end{align}

Since the minimization objective in the final formulation enforces $Z$ to have its minimum value, the value of $Z$ would be set to maximum weighted deviation from the stand-alone optimal value of each objective by \eqref{mip:cons27}. By the appropriate assignment of weights, one can prioritize the individual (or a subset of) objectives over the other objectives. For example, setting $(\gamma_1,\gamma_2,\gamma_3,\gamma_4) = (1,0,0,0)$ implies that only real power privacy preservation will be considered, while setting $(\gamma_1,\gamma_2,\gamma_3,\gamma_4) = (1,1,1,1)$ implies that all objectives will be jointly considered.

\section{Analysis}
\label{sec:Results}

This section analyzes the results of our multi-objective optimization model for different cases generated by assigning various $\gamma_i$ values. The results reveal the extent of the privacy violation when only the real power or only the reactive power is shaped as well as the improvement in privacy preservation when both real and reactive power are simultaneously shaped.

We use GAMS IDE to implement our optimization model and solve it with CPLEX on a computer with 4-core 8-thread core-i7 processor and 32GB of RAM. %As mentioned earlier, we employ the minutely real and reactive power measurement data from Ampds2 dataset~\cite{makonin2016electricity}. 

\subsection{Privacy Metric}
\label{subsec:privacymetric}

Mutual information (MI) is predominantly employed as the privacy metric in smart grid literature~\cite{erdemir2019privacy,koo2017rl,liu2017information,moon2015usages,natgunanathan2019progressive,sun2018smart}. Therefore, we adopt the same metric to measure privacy. Indeed, MI, denoted as $I(X;Y)$, is an information theoretic metric that quantifies the amount of information in one random variable $X$ that is related to another random variable $Y$. More precisely, let $p(x)$, $p(y)$, and $p(x,y)$ denote individual and joint probability distributions of $X$ and $Y$, respectively. Then, MI is defined as
\begin{equation}\label{eq:MI}
	I(X;Y) = \sum_{x \in X} \sum_{y \in Y} p(x,y) \cdot \log_{2}\frac{p(x,y)}{p(x)\cdot p(y)}.
\end{equation}

In this study, we calculate the MI between the metered real power and aggregate real power usage of all appliances as well as metered reactive power and reactive power usage of all appliances. We also calculate the MI between metered power values and individual power usage of each appliance.
%In case of real power, these two values are $p^{agg}_{t} = \sum_{a=1}^{A}p^{ca}_{a,t} + P^{sc}_t + \sum_{as}\rho_{as} \cdot P^{od}_{as,t}$ and $p^m_{t}$, respectively. Similarly, for the reactive part, the MI is calculated between
%Furthermore, we compute the MI between power usages of individual appliances and the metered power to assess the information within metered power regarding individual appliances. While calculating these MI values, both real and reactive power are considered. 
Consequently, we compare different cases by using the total MI value, which is computed as the sum of MI corresponding to real and reactive power.

Note that the resulting real and reactive power usages from the multi-objective optimization problem in this study are deterministic. Therefore, the distributions of power usages are estimated from the deterministic time series data, and the MI values are calculated according to these distributions, which is known as the \emph{empirical} MI.

\subsection{Test Bed}

For safety-critical and shiftable appliances, we choose the daily measurement data of 12/19/2012 from Ampds2, in which all appliances were used in the household. For generating the on-demand appliance usage scenarios, we used all the on-demand appliance data in the Ampds2 dataset, which accounts for 730 days of minutely appliance data. Other parameters of the computational study are given in Table~\ref{tab:parameters}.

\begin{table}[ht]
\centering \caption{Parameters and their values.}
    \renewcommand{\arraystretch}{1}
    \begin{tabulary}{7.7cm}{ccccc}
    \toprule
    Parameter && Value && Unit \\
    \midrule
    $\Delta^{t}$ && $1$ && min \\
    $\eta^{cp},\eta^{dp}$ && $0.9$ && - \\
    $\eta^{cq},\eta^{dq}$ && $0.99$ && - \\
    $E^{bi}$ && $1$ && kWh \\
    $E^{bmax}$ && $2$ && kWh \\
    $E^{ci}$ && $10$ && varh \\
    $E^{cmax}$ && $20$ && varh \\
    $P^{max}$ && $10$ && kW \\
    $R^{cbmax},R^{dbmax}$ && $0.4$ && kW \\
    $R^{ccmax},R^{dcmax}$ && $5$ && var \\
    $\epsilon$ && $10^{-3}$ && -\\
    \bottomrule
\end{tabulary}\label{tab:parameters}
\end{table}

We specified 6 cases along with the original appliance usage (case 0) for comparison to demonstrate the effectiveness of shaping both real and reactive power. Case 0 refers to the original metered load without any LS. In cases 1 and 2, only real power and only reactive power are shaped, respectively. %In case 2, privacy is preserved by only shaping the reactive power.
In case 3, real and reactive power components are shaped simultaneously. %Cases 4 and 5 represent the cases where user cost and comfort are simultaneously optimized with real power only LS and reactive power only LS, respectively.
In cases 4 and 5, cost and discomfort objectives are added to the cases with only real and only reactive power are shaped, respectively. %Case 6 represents the case where all four objectives (i.e., real power and reactive power privacy as well as the user cost and discomfort) are simultaneously optimized. 
Lastly, all objectives are jointly optimized in case 6. Table~\ref{tab:cases} summarizes all seven cases and the associated weights.

\begin{table}[ht]
\centering \caption{Cases and their corresponding weights.}
    \renewcommand{\arraystretch}{1}
    \begin{tabulary}{7.7cm}{cccccc}
    \toprule
    Case &&\multicolumn{4}{c}{Weights} \\ \cmidrule{1-1}\cmidrule{3-6}
    \# &&$\gamma_1$ & $\gamma_2$ & $\gamma_3$ & $\gamma_4$ \\
    \cmidrule{1-1}\cmidrule{3-6}
    0&& $0$ & $0$ & $0$ &  $0$ \\
    1&& $1$ & $0$ & $0$ &  $0$ \\
    2&& $0$ & $1$ & $0$ &  $0$ \\
    3&& $1$ & $1$ & $0$ &  $0$ \\
    4&& $1$ & $0$ & $1$ &  $1$ \\
    5&& $0$ & $1$ & $1$ &  $1$ \\
    6&& $1$ & $1$ & $1$ &  $1$ \\
    \bottomrule
\end{tabulary}\label{tab:cases}
\end{table}

\subsection{Effects of Shaping Real and Reactive Power on Privacy}
\label{subsec:effects1}

\begin{figure}[!b]
  \begin{center}
    \includegraphics[width=\linewidth]{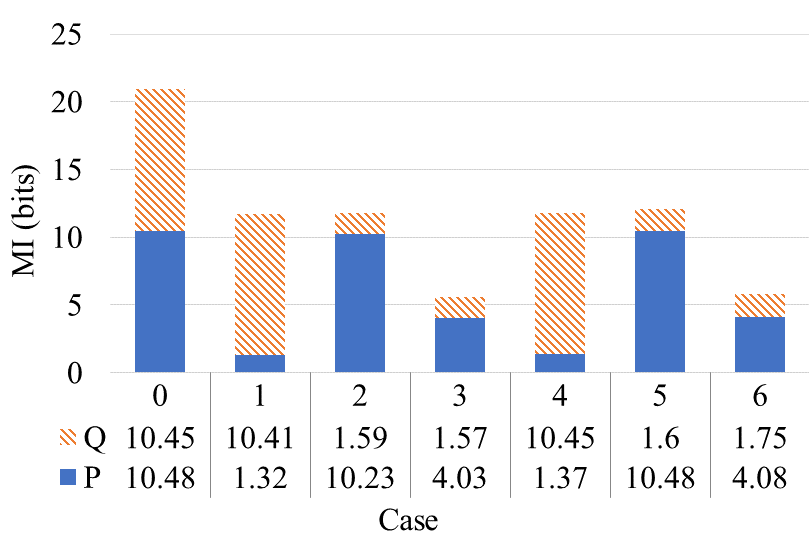}
    \caption{MI between metered and actual real ($P$) and reactive ($Q$) loads in different cases.}
    \label{fig:fig2}
  \end{center}
\end{figure}

Fig.~\ref{fig:fig2} depicts the empirical MI between actual consumption and metered load in each case for both real and reactive components. The MI of case 3 is $52\%$ less than the MIs of cases 1 and 2, where only the real power and only the reactive power are shaped, respectively. Even in case 6, where cost and discomfort objectives are also considered along with privacy objectives, there is more than a twofold increase in privacy compared to cases 1 and 2. This shows that, although shaping only real power (cases 1 \& 4) or only reactive power (case 2 \& 5) hide a significant amount of information compared to no LS (case 0), they still leak a considerable amount of information about the actual power usage, which can be further hidden by shaping real and reactive power simultaneously (cases 3 \& 6).

\iffalse
\begin{table}[!b]
\centering \caption{MI between the metered power and the actual power consumption.}
    \renewcommand{\arraystretch}{1}
    \begin{tabulary}{7.7cm}{ccccc}
    \toprule
    Case && \multicolumn{3}{c}{MI (bits)}\\
    \cmidrule{1-1}\cmidrule{3-5}
    \# && real power & reactive power & total \\
    \cmidrule{1-1}\cmidrule{3-5}
    0&& $10.48$ & $10.45$ & $20.93$ \\
    1&& $1.32$ & $10.41$ & $11.73$ \\
    2&& $10.23$ & $1.59$ & $11.82$ \\
    3&& $4.03$ & $1.57$ & $5.60$ \\
    4&& $1.37$ & $10.45$ & $11.82$ \\
    5&& $10.48$ & $1.60$ & $12.08$ \\
    6&& $4.08$ & $1.75$ & $5.83$ \\
    \bottomrule
  \end{tabulary}\label{tab:table3}
\end{table}
\fi

\begin{figure}[!htb]
  \begin{center}
    \includegraphics[width=\linewidth]{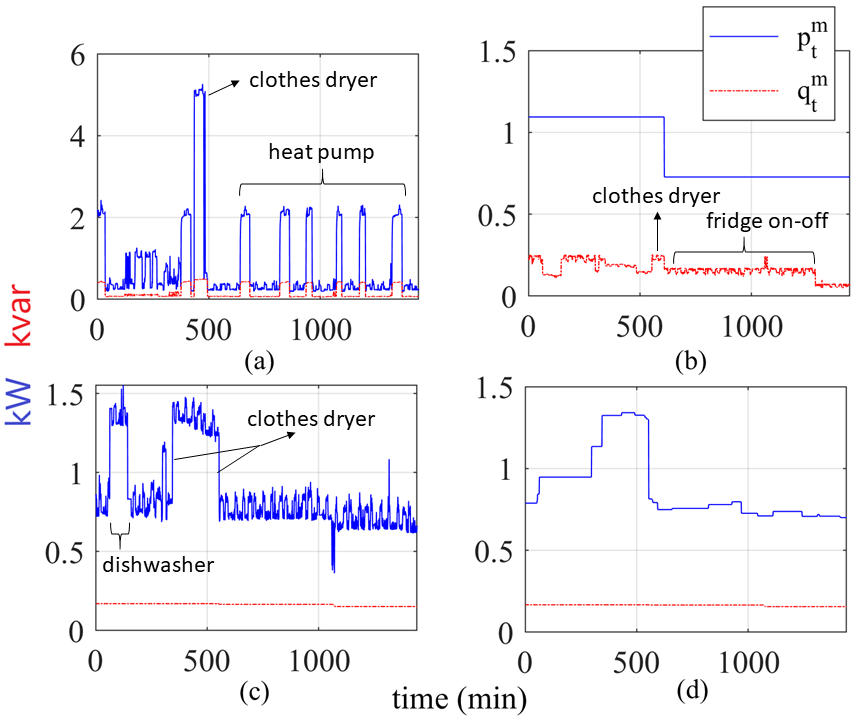}
    \caption{Metered real and reactive loads of the house. (a) original data (case 0) (b) real power based LS (case 4) (c) reactive power based LS (case 5) (d) real \& reactive power based LS optimized jointly with cost and discomfort objectives (case 6).}
    \label{fig:fig3}
  \end{center}
\end{figure}

%Fig.~\ref{fig:fig3} is helpful for the better comprehension of the temporal characteristics of the cases presented in Fig.~\ref{fig:fig2}. 
Fig.~\ref{fig:fig3} plots the metered real and reactive loads of the house in the optimal solutions of different cases. Fig.~\ref{fig:fig3}(b) and Fig.~\ref{fig:fig3}(c) show that when only real or only reactive power is shaped, the unshaped load still contains traces of many events (changes in the load), which can be mapped to appliance usages. Fig.~\ref{fig:fig3}(d) clearly illustrates that we can successfully hide most of the information on real and reactive metered loads with some deviations from their stand-alone optimal behaviors.

\begin{table}[!htb]
\centering \caption{Average MI between the metered power and the power consumed by the appliances.}
    \renewcommand{\arraystretch}{1}
    \begin{tabulary}{7.7cm}{ccccc}
    \toprule
    Case && \multicolumn{3}{c}{Average MI (bits/appliance)}\\
    \cmidrule{1-1}\cmidrule{3-5}
    \# && real power & reactive power & total \\
    \cmidrule{1-1}\cmidrule{3-5}
    0&& $3.07$ & $2.79$ & $5.86$ \\
    1&& $0.44$ & $3.29$ & $3.73$ \\
    2&& $3.47$ & $0.44$ & $3.91$ \\
    3&& $1.32$ & $0.45$ & $1.77$ \\
    4&& $0.52$ & $3.34$ & $3.86$ \\
    5&& $3.53$ & $0.47$ & $4.00$ \\
    6&& $1.33$ & $0.47$ & $1.80$ \\
    \bottomrule
  \end{tabulary}\label{tab:table4}
\end{table}
Another approach for measuring the distinguishability of the appliance footprints within the aggregate load is to calculate the MI between the measured power and the power usage of the appliances. Table~\ref{tab:table4} presents the MI between appliance loads and metered (real and reactive) loads, averaged over all appliances. This table confirms that simultaneously shaping the real and reactive loads (case 3) results in more than $52\%$ reduction in average MI compared to both real power shaping only (case 1) and reactive power shaping only (case 2). The average MI of case 6, where all objectives are considered together, is still less than half of the average MIs of cases 1 and 2 as well as the average MIs of cases 4 and 5, where cost and discomfort objectives are also considered.

\subsection{Effects of Real and Reactive LS on Other Objectives}
\label{subsec:effects2}

Quantifying the impact of shaping real and reactive power on the cost and discomfort objectives is necessary to assess the feasibility of real and reactive LS. Therefore, we analyze the effects of optimizing $O_1$ (real power privacy) and $O_2$ (reactive power privacy) on $O_3$ (cost) and $O_4$ (discomfort) in comparison to their stand-alone optimal values $O_3^\ast$ and $O_4^\ast$.

\begin{table}[!b]
\centering \caption{Effects of optimizing real and reactive power on cost and discomfort.}
    \renewcommand{\arraystretch}{1}
    \begin{tabulary}{7.7cm}{cccc}
    \toprule
    Case && \multicolumn{2}{c}{Objective value ($\%$ increase from the stand-alone optimal)}\\
    \cmidrule{1-1}\cmidrule{3-4}
    \# && cost ($O_3$) & discomfort ($O_4$) \\
    \cmidrule{1-1}\cmidrule{3-4}
    %0&& $1241$ & $2524$ \\
    4&& $46.58$ ($51.7\%$) & $2977$ ($72\%$) \\
    5&& $46.08$ ($50.1\%$) & $3237$ ($87\%$) \\
    6&& $48.48$ ($57.9\%$) & $3293$ ($90.2\%$) \\
    \bottomrule
  \end{tabulary}\label{tab:table5}
\end{table}

The results for three cases (case 4, case 5, and case 6), in which cost and discomfort weights are set to 1, are presented in Table~\ref{tab:table5}. % to observe the effects of shaping real and reactive power on the other objectives. Table~\ref{tab:table5} reveals 
It is observed that simultaneously shaping real and reactive power (case 6) increases the cost and discomfort of consumers more than shaping only real power (case 4) or shaping only reactive power (case 5) does. This is an expected behavior, as the privacy objectives are, intrinsically, conflicting with the cost and discomfort objectives. However, the rate of increase in cost in case 6 is less than $8\%$ when compared to the cost values in cases 4 and 5. This is a modest compromise when compared to more than $50\%$ decrease in the MI (i.e., more than twofold increase in privacy) as can be observed in Fig.~\ref{fig:fig2} and Table~\ref{tab:table4}. Similarly, in case 6, the increase in discomfort is less than $20\%$ when compared to the discomfort values for cases 4 and 5. 
This can be regarded as an acceptable sacrifice for a significant increase in privacy, especially when considering the fact that the discomfort objective in~\eqref{eq:o4} is a geometrically increasing function.
%(i.e., if linearly increasing penalty coefficients were adopted, the increase in discomfort could be significantly lower).

\section{Conclusion}
\label{sec:Conclusion}

%In this study, 
We presented a goal programming based multi-objective optimization framework that is capable of modeling the trade-off between LS-based real and reactive power privacy preservation, user cost, and user discomfort. We analyzed the design space by solving the optimization model for a wide range of parameters by adopting real-life data sets.
%to model appliances, ToU, and PV output. 
The major conclusions of this study are as follows:
\begin{enumerate}
\item The efficiency of privacy preservation is more than doubled (in terms of MI) when real and reactive loads are shaped simultaneously compared to the cases where only real or only reactive power are shaped. Indeed, this is the first study in the literature that investigates privacy preservation via both real and reactive power shaping in SGs and its impact on consumer cost and discomfort.
%, to the best of our knowledge.
\item The significant increase in privacy obtained by simultaneously shaping real and reactive loads comes with modest sacrifices from the cost and discomfort of consumer, which increase less than $8\%$ and less than $20\%$ from their optimal values when only real or reactive power are shaped, respectively.
\end{enumerate}

Future research directions include exploiting amenities such as batteries and PV generators for shaping real and reactive load simultaneously without the need for a household capacitor. Although currently PV generators are restricted to have a unity power factor, they can, potentially, be used for reactive power compensation~\cite{kekatos2015stochastic} which can also help to shape the reactive load for privacy.
%We present the optimal privacy results in a demand response scheme by shaping both real and reactive metered power. Our results reveal that shaping only the real or the reactive load can result in serious data leakage for consumers. Major takeaways from this study are as follows:
%\begin{enumerate}
%\item When real power ($P$) and reactive power ($Q$) are shaped together, the optimal privacy is enhanced more than $50\%$ in terms of mutual information. In other words, when $P$ or $Q$ is shaped alone, the total mutual information between the actual and metered loads increase at least twofold compared to shaping $P$ and $Q$ together.
%\item Optimizing privacy by shaping both the real and reactive power results in $57.9\%$ and $90.2\%$ increase in consumer cost and discomfort, respectively, from their singular optimal values. However, the additional cost of shaping both $P$ and $Q$ is less than $8\%$ and the additional discomfort is less than $20\%$ when compared to real-power-only or reactive-power-only privacy optimization, respectively.
%\end{enumerate}
%Future research directions include using amenities such as batteries and PV generators for shaping the real and the reactive load simultaneously, without the need for a household capacitor. Although currently PV generators are restricted to have unity power factor, they can be used for reactive power compensation \cite{kekatos2015stochastic} which can also help shaping the reactive load for privacy.

\bibliographystyle{IEEEtranTIE}
\bibliography{pq-opt-v007}

%\begin{IEEEbiography}[{\includegraphics[width=1in,height=1.25in]{Author_Cihan_Emre-Kement.pdf}}]{Cihan Emre Kement}
%\end{IEEEbiography}
%%\vspace{-10mm}
%\begin{IEEEbiography}[{\includegraphics[width=1in,height=1.25in]{Author_Hakan_Gultekin.pdf}}]{Hakan Gultekin}
%\end{IEEEbiography}
%%\vspace{-10mm}

\begin{IEEEbiography}[{\includegraphics[width=1in,height=1.25in]{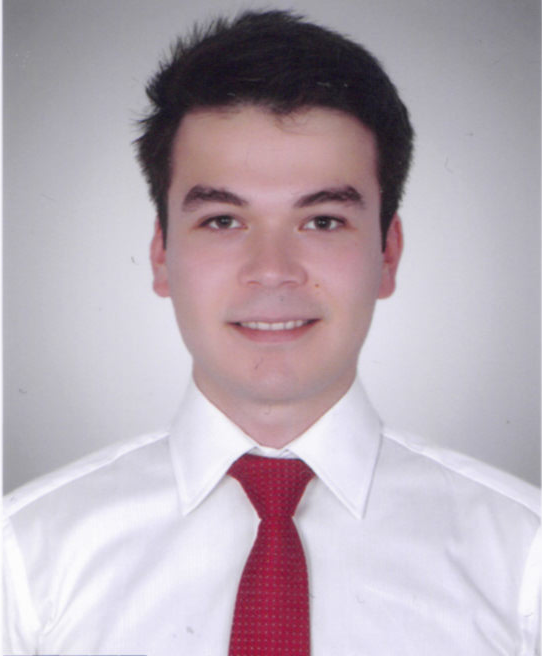}}]{Cihan Emre Kement} [M] (ckement[at]etu.edu.tr) received the B.Sc., M.Sc., and Ph.D. degrees in electrical and electronics engineering from Bilkent University, Middle East Technical University, and TOBB University of Economics and Technology, Ankara Turkey in 2011, 2014, and 2020, respectively. He was a Fulbright visiting researcher at Laboratory for Information and Decision Systems (LIDS), Massachusetts Institute of Technology (MIT), Cambridge, MA, USA.
His current research interests are security and privacy in cyber-physical systems, wireless communications and optimization.
\end{IEEEbiography}
\vspace{-1cm}

\begin{IEEEbiography}[{\includegraphics[width=1in,height=1.25in]{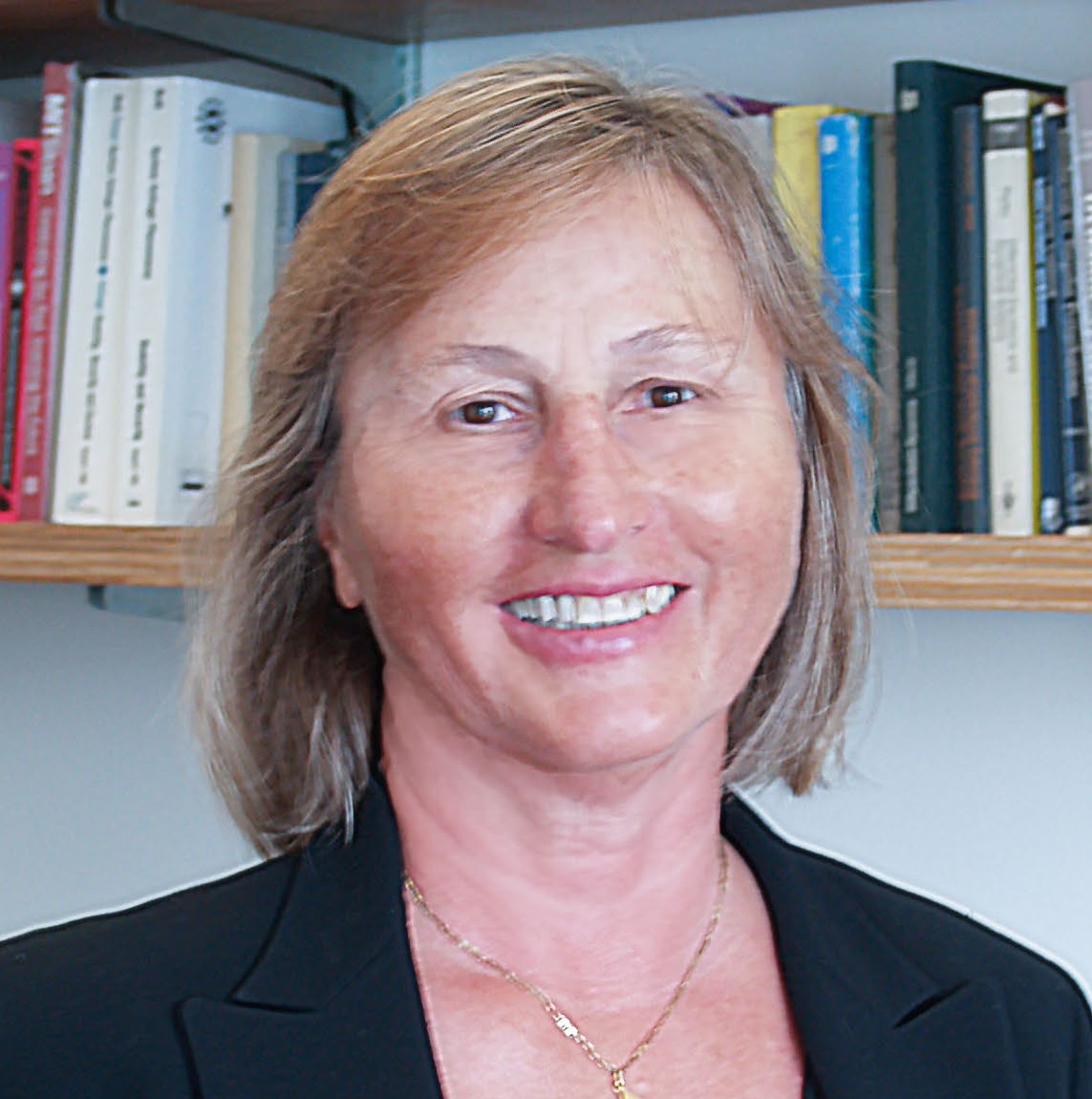}}]{Marija Ili\'c} [F] (ilic[at]mit.edu) is a Senior Research Scientist at MIT LIDS Laboratory, and Professor Emerita of Electrical \& Computer Engineering and Engineering \& Public Policy at Carnegie Mellon University. She was an Assistant Professor at Cornell University, Ithaca, NY, and tenured Associate Professor at the University of Illinois at Urbana-Champaign. She was a Senior Research Scientist in Department of Electrical Engineering and Computer Science at MIT from 1987 to 2002. She has over 30 years of experience in teaching and research in the area of electrical power system modeling and control. Her main interest is in the systems aspects of operations, planning, and economics of the electric power industry. She has co-authored several books in her field of interest. She is an IEEE Fellow and Distinguished Lecturer.
\end{IEEEbiography}
\vspace{-1cm}

\begin{IEEEbiography}[{\includegraphics[width=1in,height=1.25in]{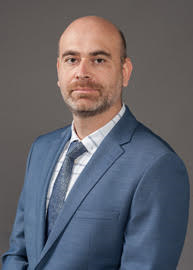}}]{Hakan Gultekin} (hgultekin[at]squ.edu.om) 
received the B.Sc., M.Sc., and Ph.D. degrees in industrial engineering from Bilkent University, Ankara, Turkey, in 2000, 2002, and 2007, respectively. He has been an Associate Professor with the Department of Mechanical and
Industrial Engineering, Sultan Qaboos University, Muscat, Oman, since September 2018. He has also been affiliated with the Department of Industrial Engineering, TOBB University of Economics and Technology, Ankara, since 2007. His research interests include scheduling, optimization modeling, and exact and heuristic algorithm development, especially for problems arising in communication systems, modern manufacturing systems, energy systems, and wireless sensor networks.
\end{IEEEbiography}
\vspace{-1cm}

\begin{IEEEbiography}[{\includegraphics[width=1in,height=1.25in]{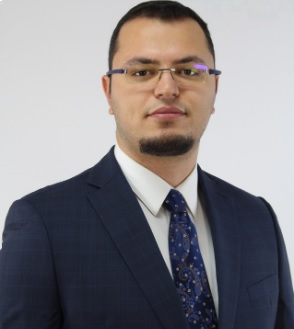}}]{Cihan Tugrul Cicek} (cihan.cicek[at]atilim.edu.tr) 
received the B.Sc. degree in industrial engineering from TOBB University of Economics and Technology, Ankara, Turkey, in 2010; M.Sc. degrees in operations research from Middle East Technical University, Ankara, Turkey and in facilities and environmental management from the University College London, London, U.K., in 2014; and the Ph.D. degree in industrial engineering from TOBB University of Economics and Technology, Ankara, Turkey, in 2019. He is currently an Assistant Professor with the Department of Industrial Engineering, Atilim University, Ankara, Turkey. His research interests include mathematical optimization and algorithms with applications in wireless communications, aerial networks, facility location and smart grids.
\end{IEEEbiography}
\vspace{-1cm}

\begin{IEEEbiography}[{\includegraphics[width=1in,height=1.25in]{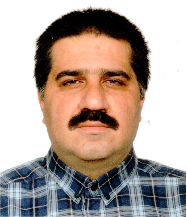}}]{Bulent Tavli} [SM] (btavli[at]etu.edu.tr) received his Ph.D. degree in electrical engineering from the University of Rochester, Rochester, NY, USA, in 2005. He is, currently, a professor with the department of electrical and electronics engineering, TOBB University of Economics and Technology, Ankara, Turkey. His research interests include network science, telecommunications, optimization, machine learning, information security and privacy, smart grids, embedded systems, and blockchain.
\end{IEEEbiography}

\end{document}